\documentclass[epjCONF,columns]{svjour} 
\usepackage{graphics} \usepackage[varg]{txfonts} 
\usepackage[latin1]{inputenc}
\session-title{2011 Hadron Collider Physics symposium (HCP-2011)}
\begin{document}

\hyphenation{pha-ses} \hyphenation{ha-dro-nic}
\hyphenation{ca-lo-ri-me-ter} \hyphenation{to-ro-id-al}
\hyphenation{se-arch-ing} \hyphenation{spec-tros-copy}
\hyphenation{pro-pos-ed} \hyphenation{track-ing}

\title{ATLAS Upgrade for the HL-LHC: meeting the challenges of a
  five-fold increase in collision rate}

\author{Peter
  Vankov\inst{1}\fnmsep\thanks{\email{peter.vankov@cern.ch}}}
\institute{Deutsches Elektronen Synchrotron, DESY, Notkestrasse 85,
  22607 Hamburg, Germany
  ~\\
  ~\\
  \emph{On behalf of the ATLAS collaboration} }
\abstract{With the LHC successfully collecting data at $7$~TeV, plans
  are actively advancing for a series of upgrades leading eventually
  to about five times the LHC design-luminosity some $10$-years from
  now in the High-Luminosity LHC (HL-LHC) project.  Coping with the
  high instantaneous and integrated luminosity will require many
  changes to the ATLAS detector.  The designs are developing rapidly
  for an all-new inner-tracker, significant changes in the calorimeter
  and muon systems, as well as improved triggers.  This article
  summarizes the environment expected at the HL-LHC and the status of
  various improvements to the ATLAS detector.} 
\maketitle
\section{Introduction}
\label{intro}
ATLAS~\cite{atlas_tdr}, at the CERN Large Hadron Collider
(LHC)~\cite{lhc_tdr}, is a general-purpose experiment designed to
explore the proton-proton (\emph{pp}) collisions at the LHC with
center-of-mass energies of up to $\sqrt{s}=14$~TeV at a maximum
luminosity of $\mathcal{L}^{\textrm{peak}}=10^{34}$~cm$^{-2}$s$^{-1}$.
The high collision energy along with the high luminosity at the LHC
would eventually allow observation of \emph{new physics} at the TeV
scale.  ATLAS is constructed to fully exploit the physics potential of
the LHC, which includes the discovery of the Higgs particle, as well
as searches for effects beyond the Standard Model (SM).

\begin{figure}[!h]
  \centering \resizebox{1.\columnwidth}{!}{%
    \includegraphics{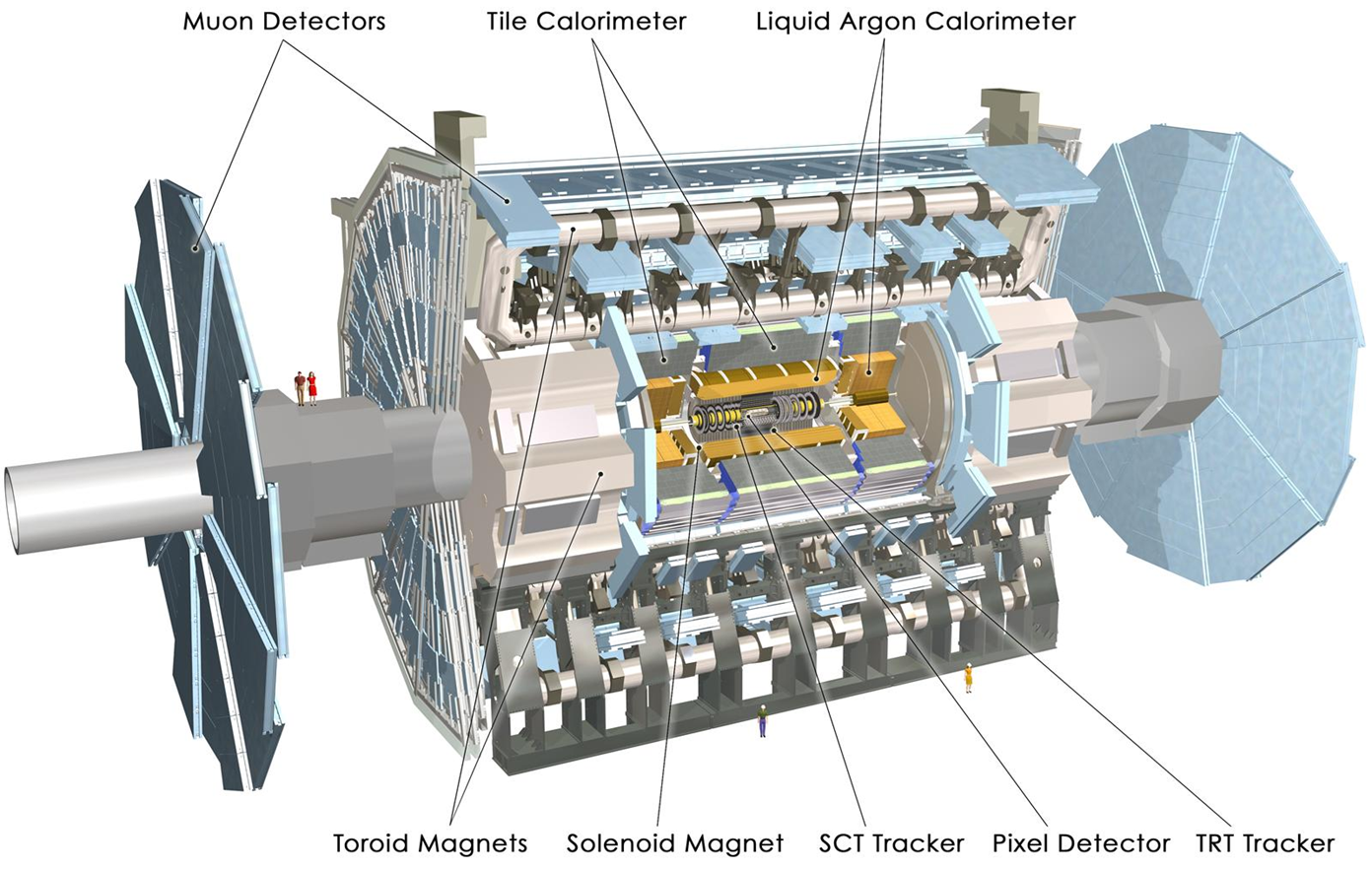} }
  \caption{The ATLAS experiment.}
  \label{atlas}
\end{figure}

As illustrated in Fig.~\ref{atlas}, ATLAS comprises three basic
subsystems: the Inner Detector, ID (Pixel, SCT, TRT), housed in a
solenoid creating a magnetic field of $2$~T, the Calorimetry system
(Liquid Argon and Tile) and the Muon Spectrometer (MS) with its
associated superconducting toroidal magnets supplying a magnetic field
of $0.5$~T.
A three-level trigger system is used to select the events of interest,
providing a final average trigger-rate of a few hundred Hz.  The
overall dimensions of $44$~m in length and $25$~m in diameter, make
ATLAS
the largest detector in collider experiments.

ATLAS has been successfully taking \emph{pp} collision data at
$\sqrt{s}=7$~TeV ($3.5$~TeV per beam) since March $2010$.  As a result
of the excellent performance and operation of the experiment, by the
end of October $2011$, ATLAS has recorded an integrated luminosity of
$5$~fb$^{-1}$ with stable beams, corresponding to an overall
data-taking efficiency of $94$\%.

In the next years, LHC will undergo a series of upgrades leading
ultimately to five times increase of the instantaneous luminosity in
the High-Luminosity LHC (HL-LHC) project.  The goal is to extend the
dataset from about $300$ fb$^{-1}$, expected to be collected by the
end of the LHC run (in $\sim 2020$), to $3000$~fb$^{-1}$ by $\sim
2030$. The foreseen higher luminosity at the HL-LHC is a great
challenge for ATLAS. Meeting it will require significant detector
optimizations, changes and improvements, which are subject of these
proceedings.

\section{HL-LHC and the ATLAS Upgrade Plans}
\label{hl-lhc_and_atlas_upgrade_plans}

The main motivation for the HL-LHC is to extend and to improve the LHC
physics program \cite{kjacobs_slhc_physics}.
Depending on the results from the LHC data, some of the physics
problems that could be addressed at the HL-LHC are: measuring of the
Higgs rare decays and Higgs self-couplings; performing a complete
supersymmetry spectroscopy; searching (extending limits) for new gauge
bosons ($W^{\prime}, Z^{\prime}$); searching for a quark and lepton
substructure.

The harsher radiation environment and higher detector occupancies at
the HL-LHC imply major changes to most of the ATLAS systems, specially
those at low radii and large pseudorapidity, $\eta$. A general
guideline for these changes is maintaining the same (or better) level
of detector performance as at the LHC. The ID, forward calorimeter and
forward muon wheels will be affected the most by the higher particle
fluxes and radiation damage, requiring replacement or significant
upgrade, whereas the barrel calorimeters and muon chambers are
expected to be capable of handling the conditions and will not be
modified. New, radiation-hard tracking detectors with higher
granularity and higher bandwidth, as well as radiation-hard front-end
(FE) electronics are foreseen.  The higher event rates and event sizes
will be a challenge for the trigger and data acquisition (DAQ)
systems, which will require a significant expansion of their capacity.

The ATLAS upgrade is planned in three phases, which correspond to the
three long, technical shutdowns of the LHC towards the HL-LHC.
\emph{Phase-0} ($\sim24$ months) will take place in $2013$ and $2014$,
the \emph{Phase-I} ($\sim12$ months) will be during $2018$, and
finally, the \emph{Phase-II} ($\sim24$ months) is scheduled for
$2022$-$2023$.
\section{ATLAS Upgrade: Phase 0}
\label{sec:phase_0}

The main objective of the $2013$-$2014$ shutdown of the LHC is to
perform interventions which will permit the machine to operate at its
design parameters: center-of-mass energy of $\sqrt{s}=14$~TeV and
luminosity of $1 \times 10^{34}$~cm$^{-2}$s$^{-1}$.

ATLAS will use this two years period for detector consolidation works,
including a new ID cooling system, a new neutron shielding of the MS,
and a new beam pipe. The current beam pipe in the forward region is
made of stainless steel which is a source of high backgrounds for the
MS. The new beam pipe will be of aluminum, thus, reducing the
backgrounds by $10$-$20$\%.

The central ATLAS upgrade activity in Phase-0 is the installation of a
new barrel layer in the present Pixel detector, the so called IBL
project.
\subsection{Insertable B-layer}
\label{sec:phase_0-ibl}

\begin{figure}[!h]
  \centering \resizebox{0.6\columnwidth}{!}{%
    \includegraphics{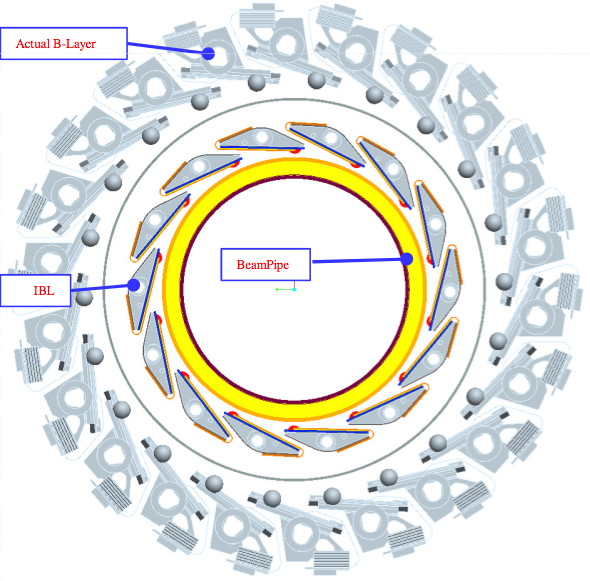} }
  \caption{Cross-section view of the current Pixel B-layer, the new
    beam pipe and the IBL.}
  \label{ibl}
\end{figure}

The Insertable B-layer (IBL) is an additional, $4^{\textrm{th}}$ pixel
layer which will be inserted between the innermost pixel layer (the
B-layer) and the beam pipe, as shown in Fig.~\ref{ibl}, during the
Phase-0 upgrade. To make the installation of the IBL possible, a new
beam pipe in the central region, with reduced by $4$~mm radius
(r=$29$~mm $\rightarrow$ r=$25$~mm), built of Beryllium, is envisaged.

As demonstrated in Ref~\cite{ibl_tdr}, it is expected that the IBL
will improve the vertex resolution, secondary vertex finding and
b-tagging, hence extending the reach of the physics analysis. It will
compensate for defects (irreparable failures of modules) in the
existing B-layer, assuring tracking robustness.  Moreover, IBL will
help to preserve the tracking performance at the luminosity beyond
$\mathcal{L}^{\textrm{peak}}$, e.g. in Phase-1, when the B-layer will
suffer from radiation damage and high pile-up occupancies.

The baseline concept of the IBL consists of $14$ staves, mounted
directly on the beam pipe with a tilt angle of $14^{\circ}$.  On each
stave there are $16$ to $32$ modules depending on the sensor
type. Currently, two silicon sensor types are under consideration:
planar and 3D.
The IBL modules will be equipped with a new readout chip,
FE-I4~\cite{fe-i4}, which has been specially developed to function at
high data transfer rates ($\sim 160$~Mb/s). The FE-I4 design allows an
increase of the IBL segmentation by decreasing the pixel size from
$50$~$\mu$m $\times$ $400$~$\mu$m to $50$~$\mu$m $\times$
$250$~$\mu$m.
\section{ATLAS Upgrade: Phase-I}
\label{phase_1}

In $2018$, the LHC will be stopped for an upgrade of the injectors and
the collimators.  Upgrade of the LINAC2 and increase of the Proton
Synchrotron Booster output energy are planned.  The data-taking will
be resumed after one year shutdown with luminosity of $2 \times
10^{34}$ cm$^{-2}$s$^{-1}$. During the shutdown, ATLAS intends to
accomplish the second stage of its upgrade program, the Phase-I.

In Phase-I, installation of new Muon Small Wheels and introducing of
new trigger schemes (Fast TracKer, topological triggers, improved
L1Calo granularity) are proposed to handle luminosities well beyond
the nominal values.

\subsection{New Muon Small Wheels}
\label{sec:new_muon_small_wheels}

A replacement of the first endcap station of the Muon Spectrometer,
the Muon Small Wheel (MSW), built of Monitored Drift Tubes (MDT) and
Cathode Strip Chambers (CSC), is proposed. The concern is that for
luminosities $\mathcal{L} > \mathcal{L}^{\textrm{peak}}$, in addition
to the higher number of pile-up events per bunch-crossing, large
amounts of cavern background will be induced, affecting a large
$|\eta|$ region of the MSW. The current system in this region will
struggle badly to cope with this and therefore a replacement is
required.

The new Muon Small Wheels must ensure efficient tracking at high
particle rate (up to $\mathcal{L}=5 \times 10^{34}$ cm$^{-2}$s$^{-1}$
) and large $|\eta|$, with position resolution of $< 100$~$\mu$m.
Furthermore, the new MSW will be integrated into the Level-1 trigger
\cite{muon_small_wheel}. Several detector technologies are under
investigation at the moment: small diameter MDT's (sMDT) complemented
with fast trigger chambers - Resistive Plate Chambers (RPC) or Thin
Gap Chambers (TGC); Fine strip TGC's; Micro-MEsh GAseous Structures
(MicroMEGAs); or some other combinations of these.
\subsection{New Trigger Schemes}
\label{sec:new_trigger_schemes}

At Phase-I, more sophisticated triggers will be required.  For this,
the Fast TracKer (FTK) trigger project has been initiated \cite{ftk}.
At the FTK, the track finding and fitting are conducted at a hardware
level, which makes it extremely fast. At the current ATLAS, this task
is performed by the trigger Level-2 software farm.  FTK will provide
the track parameters at the beginning of the Level-2 processing. This
way, the load on Level-2 will be diminished and extra resources will
be available for more advanced selection algorithms, which ultimately
could improve the b-tagging, lepton identification, etc.

Suggestions are also in place for combining trigger objects at Level-1
(topological triggers) and for implementing full granularity readout
of the calorimeter. The latter will strongly improve the triggering
capabilities for electrons and photons at Level-1.

\section{ATLAS Upgrade: Phase-II}
\label{phase_2}

The ATLAS Phase-II upgrade is scheduled for $2022$ and $2023$. During
this time, LHC will be out of operation for furnishing with new inner
triplets and crab cavities. As a result, an instantaneous luminosity
of $5 \times 10^{34}$ cm$^{-2}$s$^{-1}$ should be achieved. The goal
is to accumulate $3000$ fb$^{-1}$ of data by $\sim 2030$.

ATLAS Phase-II preparations include a new Inner Detector and further
trigger and calorimeter upgrades.

\subsection{New Inner Detector}
\label{sec:itk}

Running at nominal $\mathcal{L}^{\textrm{peak}}$ for the LHC , will
bring, on average, $\sim 28$ primary interactions (pile-up events) per
bunch crossing, every $25$ ns. The number of pile-up events at
$5\times10^{34}$ cm$^{-2}$s$^{-1}$ is therefore expected to be $\sim
140$. (Should luminosity levelling not be fully effective or some
other scheme adopted, $7\times 10^{34}$ cm$^{-2}$s$^{-1}$ should at
least be accommodated.) This will result in $5$ to $10$ times higher
detector occupancies, which is beyond the TRT design parameters.
Furthermore, by $2022$, the Pixel and the SCT subsystems, would
seriously degrade their performance due to the radiation damage of
their sensors and FE electronics. Because of all these factors, ATLAS
has decided to replace the entire Inner Detector with a new,
all-silicon \emph{Inner Tracker} (ITk). The ITk must satisfy the
following criteria (w.r.t. ID): higher granularity, improved material
budget, increased radiation resistivity of the readout components.  At
the moment, the ITk project is in an R\&D phase.  Different
geometrical layouts are simulated and their performance is studied in
search for the optimal tracker architecture.  A major constraint on
the design is the available space, defined by the volume taken by the
ID in ATLAS. This implies a maximum radius of $\sim 1$~m and the
limiting existing gaps for services.

\begin{figure}[!h]
  \centering \resizebox{0.48\columnwidth}{!}{%
    \includegraphics{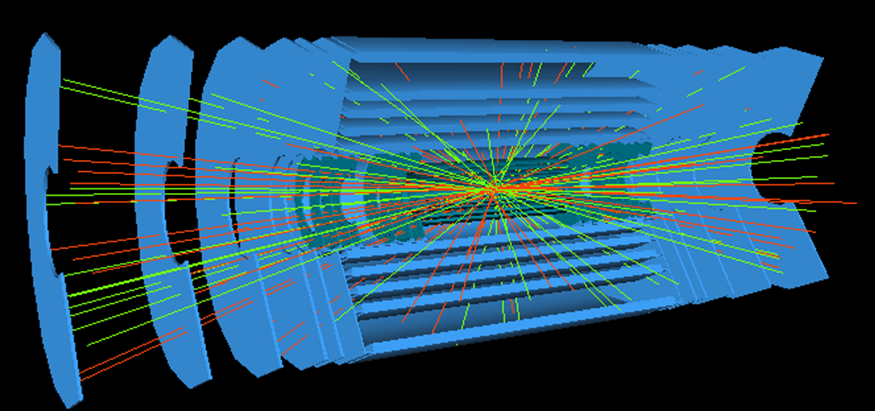} } \centering
  \resizebox{0.48\columnwidth}{!}{%
    \includegraphics{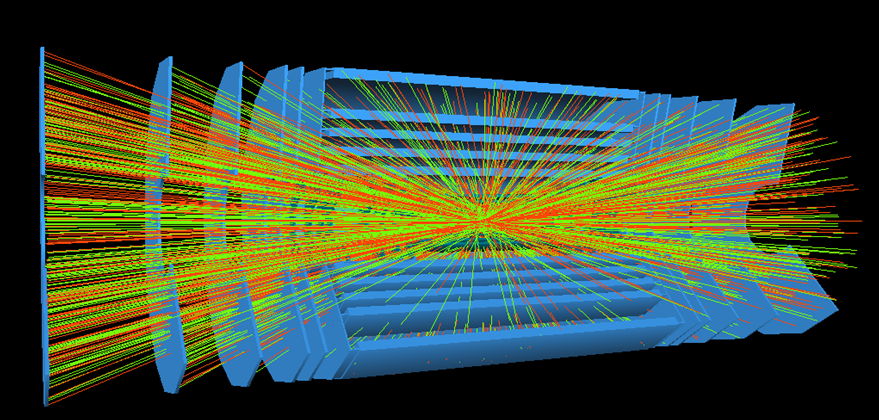} }
  \caption{The baseline layout of the new Inner Detector, traversed by
    simulated 23 pile-up events (left) and 230 pile-up events
    (right).}
  \label{utopia}
\end{figure}

The current baseline design of the ITk, depicted in Fig. \ref{utopia},
consists of $4$ Pixel and $5$ Si-strip layers in the barrel part. The
two endcap regions are each composed of $6$ Pixel and $5$ Si-strip
double-sided disks, built of rings of modules. The pixel modules are
with identical pixels of size $50\times250$ $\mu$m, whereas the
Si-strip modules come in two types, with short ($24$~mm) and long
($96$~mm) strips. As in the current SCT, the Si-strip modules are
designed to be of $2$ pairs of silicon microstrip sensors, glued
back-to-back at an angle of $40$ mrad to provide 2D space-points.

Intensive R\&D studies are also in process to select the most suitable
pixel sensor technology out of Si-planar, 3D and diamond, and to find
the optimal layout of the Si-strip modules \cite{si_microstips}.

\subsection{Calorimeter and trigger upgrades}
\label{sec:phase2_calo_trig}

The HL-LHC conditions will have a major impact on the Calorimetry
system.  To ensure an adequate performance, a replacement of the cold
electronics inside the LAr Hadronic endcap, as well as, a replacement
of all on-detector readout electronics for all calorimeters may need
to be anticipated. Also, the operation of the Forward Calorimeter
(FCal) could be compromised. To maintain the FCal functioning at the
HL-LHC, two possible solutions are considered \cite{fcal}: first,
complete replacement of the FCal, and second, installation of a small
warm calorimeter, Mini-FCal, in front of the FCal. The Mini-Fcal would
reduce the ionization and heat loads of the FCal to acceptable levels.

The planned trigger upgrades for Phase-II, are connected with
implementing a Track Trigger at Level-1/Level-2, applying full
granularity of calorimeter at Level-1 and improving the muon trigger
coverage.

\section{Conclusions}
\label{conclusions}

ATLAS collaboration has devised a detailed program to reflect the
changes in the LHC conditions towards the High-Luminosity LHC,
characterized by high track multiplicity and extreme fluences. At each
of the 3 phases of the upgrade program, actions will be undertaken to
reassure the stable and efficient performance of the ATLAS detector.

\end{document}